\documentclass{article}
 \usepackage{graphicx}

\def\beq{\begin{equation}}
\def\eeq{\end{equation}}
\def\bea{\begin{eqnarray}}
\def\eea{\end{eqnarray}}
 
\begin{document}
 
\title{\bf Collective motion in finite Fermi systems within Vlasov dynamics}
\author{ V. I. Abrosimov$^{\rm a}$, A.Dellafiore $^{\rm b}$,
F.Matera$^{\rm b}$\\
\it $^{\rm a}$ Institute for Nuclear Research, 03028 Kiev, Ukraine\\
$^{\rm b}$ \it Istituto Nazionale di Fisica Nucleare\\
\it  and Dipartimento di Fisica, Universit\`a degli Studi di Firenze,\\
\it via Sansone 1, I-50019 Sesto F.no (Firenze), Italy}
\date{}
\maketitle
\begin{abstract}
A semiclassical  theory of linear response in finite Fermi systems, based on the Vlasov equation, and its applications to the study of isoscalar vibrations in heavy nuclei are reviewed. It is argued that the Vlasov equation can be used to study the response of small quantum systems like (heavy) nuclei in regimes for  which the finite size of the system is more important than the collisions between constituents. This requires solving the linearized Vlasov equation for finite systems, however, in this case the problem of choosing appropriate boundary conditions for the fluctuations of the phase-space-density is non-trivial. Calculations of the isoscalar response functions performed by using different boundary conditions, corresponding to fixed and moving nuclear surface, are compared for different multipoles and it is found that, in a sharp-surface model, the moving-surface boundary conditions give better agreement with experiment. The semiclassical strength functions given by this theory are strikingly similar to the results of analogous quantum calculations, in spite of the fact that shell effects are not included in the theory, this happens because of a well known close
relation between classical trajectories and shell structure. 
\end{abstract}

\section{Introduction}
\bigskip
\bigskip

        The Landau kinetic equation for Fermi liquids \cite{lan1,lan2} contains some important 
differences when compared to the classical Boltzmann  equation for a dilute gas, one of them is 
the presence of an effecive mean-field term. Thus, in Landau's approach, at least part of the 
force which is exerted on a particle by the other constituents of a many-body system  can be 
approximated by an effective mean field.  Another important feature of the Landau kinetic equation
 is the introduction of an effective mass for quasiparticles, however here we neglect the difference between bare and effective  mass. Landau's kinetic equation allows also for a collision term which, nonetheless, in some cases may be neglected (see for example Sect. 4 of \cite{lipi9}). When this is done, we are dealing with a collisionless kinetic equation in the mean field approximation. Such 
an  approximation had been considered long ago by Jeans (who gave the credit to Boltzmann)  in 
the context of stellar dynamics \cite{jea} and later by Vlasov in that of plasma physics 
\cite{vla} . Here we follow the use that has become common both in plasma and  nuclear physics 
and refer to the collisionless kinetic equation in the mean field approximation as the Vlasov 
equation (see \cite{hen} for a discussion of historical priorities).

        Kirzhnitz and collaborators \cite{kir} extended the approach to non homogeneous systems 
and used it to study the possibility of collective excitations in the electron cloud of heavy 
atoms. They  pointed out that a main difficulty arising in finite systems concerns the 
boundary conditions to be imposed on the fluctuations of the phase-space-density. Moreover these
authors also derived an interesting expression for the polarization propagator determining the 
linear response of these systems. Unfortunately the practical usefulness of this expression is  limited
to rather special systems in which the constituents move along closed orbits.

        Another attempt to study the dynamic response of inhomogeneous Fermi systems was  limited to one-dimensional problems \cite{cha}.

        Bertsch \cite{ber} argued that the Vlasov equation could be used as a starting point 
for a semiclassical theory of giant resonances in heavy nuclei. He pointed out that, in spite
 of being a classical equation of motion, this equation would not violate the Pauli principle,
 at least in a semiclassical sense. This is a consequence of the Liouville theorem. When applying
 this method to nuclei, however, one is faced with the problem of finite-size effects since a 
nucleon close to the Fermi surface is more likely to reach the nuclear surface than to suffer 
a violent collision with another nucleon. Therefore finite-size effects become more important 
than the collision integral and also in the case of nuclei it is reasonable to study the kinetic
 equation in the mean-field (or Vlasov) approximation, at least as a first step.

        Some remarkable progress on this problem has been made
        in the field of galactic dynamics: Polyacenko and Shukhman \cite{pol} solved the 
linearized Vlasov equation for finite spherical systems in their study of the stability of 
collisionless  stellar systems. In this context one of the main problems is that of determining 
a stable equilibrium distribution of particles (stars). Small deviations from the equilibrium 
distribution are characterized by eigenfrequencies that are purely imaginary in the case of unstable 
systems. Their approach has found several applications  in the field \cite{ber2,pal,ber3}.

        A similar solution of the linearized Vlasov equation for nuclear response has been derived
 independently in \cite{bri}. This solution turns out to agree with that of \cite{pol} and gives a reasonable description of giant resonances in heavy nuclei \cite{dit}.

        Abrosimov, Di Toro and Strutinsky \cite{ads} used the same approach within a sharp-surface model in which the nuclear mean field is approximated by a square-well potential and used also different (moving-surface) boundary conditions in order to extend the approach of \cite{bri} to low-energy surface vibrations in heavy nuclei.

        This paper is a review of work done in the last ten years, based on the approaches 
of \cite{bri} and \cite{ads}.  In Sect. 2 both  approaches are recalled, while in Sect. 3 several applications of the theory to the study of isoscalar vibrations of different multipolarity (monopole,  dipole,  quadrupole,  octupole) in heavy nuclei are discussed. Finally, in Sect. 4, conclusions are drawn. The two Appendices contain some more technical material on the moving-surface response functions and on the Fourier coefficients that replace the quantum matrix elements in our semiclassical approach.

\section{Reminder of formalism}
\subsection{Smooth surface}

        In our semiclassical approach we assume that a (heavy) spherical nucleus in its ground 
state can be described  by the following equilibrium phase-space-distribution\footnote{We use units $\hbar=c=1$}
        \beq
        \label{eqd}
        f_0({\bf r},{\bf p})=\frac{4}{(2\pi\hbar)^3}\vartheta(\epsilon_F-h_0({\bf r},
{\bf p}))=F(h_0)\,,
        \eeq
where $\vartheta$ is the step function,  $\epsilon_F$ is the Fermi energy, while
\beq
\label{qpe}
h_0({\bf r},{\bf p})=\frac{p^2}{2m}+V_0(r)=\epsilon
\eeq
is the quasiparticle energy
and the equilibrium mean field $V_0(r)$ is assumed to be spherical. In principle the mean field 
should be determined self-consistently as (Hartree approximation)
\beq
\label{eqmf}
V_0({\bf r})=\int d{\bf r}' v({\bf r},{\bf r'})\varrho_0({\bf r}')\,,
\eeq
where $v({\bf r},{\bf r'})$ is the effective interaction between quasiparticles and
\beq
\label{eqden}
\varrho_0({\bf r})=\int d{\bf p}f_0({\bf r},{\bf p})
\eeq
the equilibrium density of the nucleus (for sake of simplicity we do not take into account 
explicitly the spin and isospin degrees of freedom since this would only complicate the formalism
 without posing any new conceptual difficulty; the statistical weight $4$ in Eq.(\ref{eqd}) 
accounts for these degrees of freedom). In practice we shall use instead a phenomenological 
equilibrium mean field, which can be either of the Saxon-Woods shape or even a simple square-well
 potential of radius $R=1.2 A^{\frac{1}{3}}{\rm fm}$. For the square-well approximation, several 
expressions can be evaluated analytically and this is one of the merits of our simplified approach. Note that, contrary to the standard Fermi liquid theory, our equilibrium distribution (\ref{eqd}) depends also on the space coordinate ${\bf r}$.

        Next we assume that at time $t=0$ the system is subject to an external driving field of 
the kind
\beq
\label{vext}
V^{ext}({\bf r},t)=\beta\delta(t)Q({\bf r})\,.
\eeq
Here $\beta$ is a parameter determining the intensity of the external force, which is applied only 
for a very short time around $t=0$, as described by the Dirac $\delta$-function, and $Q({\bf r})$ 
gives the space dependence of the external field. Typically we shall be interested in the multipole 
response of order $L$, for which\footnote{ For compression modes we'll be interested also in $Q({\bf r})=r^{L+2} Y_{LM}({\bf \hat r})$.}
\beq
\label{mul}
Q({\bf r})=r^L Y_{LM}({\bf \hat r}).
\eeq
        The response of the system to an external force is described by the fluctuation of the 
phase-space density defined by
        \beq
        \label{flu}
        f({\bf r},{\bf p},t)=f_0({\bf r},{\bf p})+\delta f({\bf r},{\bf p},t)
        \eeq
or, equivalently, by its time Fourier transform
\beq
\label{fft}
\delta f({\bf r},{\bf p},\omega)=\int_{-\infty}^{\infty} dt e^{i\omega t} \delta f({\bf r},{\bf p},
t)\,.
\eeq
Since $\delta f({\bf r},{\bf p},t)$ vanishes for $t<0$, we can suppose that $\omega$ has a 
vanishingly small imaginary part $i\varepsilon$ to ensure the convergence of this integral 
when $t\to +\infty$.

        The perturbed system is described by a time-dependent phase-space-density satisfying the 
mean-field (or Vlasov) kinetic equation
\beq
\label{vkin}
\frac{\partial f}{\partial t}+\{f,h\}=0\,,
\eeq
where the braces are Poisson brackets.
The time-dependent Hamiltonian $h$ is given by
\beq
\label{tdh}
h({\bf r},{\bf p},t)=h_0({\bf r},{\bf p})+\delta h({\bf r},t)\,,
\eeq
with
\beq
\label{delh}
\delta h({\bf r},t)=V^{ext}({\bf r},t)+\delta V^{int}({\bf r},t)\,.
\eeq

This expression shows explicitly that the extra force acting on a particle in the perturbed system 
has two components: one due to the external driving field $V^{ext}$ and an additional one  due to 
the change in the interaction with the surrounding particles. This last term is given by
 \beq
 \label{dvint}
 \delta V^{int}({\bf r},t)=\int d{\bf r}' v({\bf r},{\bf r'})\int d{\bf p}'\delta f({\bf r}',
{\bf p}',t)\,.
 \eeq

        If the external force is sufficiently weak, the density fluctuation induced by it is small 
and we can consider only terms that are linear in $\beta$. In this case the fluctuation $\delta 
f({\bf r},{\bf p},t)$ satisfies the linearized Vlasov equation
        \beq
        \label{lve1}
        \frac{\partial\,\delta f}{\partial t}+\{\delta f,h_0\}+\{f_0,\delta h\}=0\,,
        \eeq
        or
        \beq
        \label{lve2}
        \frac{\partial\,\delta f}{\partial t}+\{\delta f,h_0\}=-F'(h_0)\{h_0,\delta h\}\,.
        \eeq

        From the mathematical point of view, if the $({\bf r}, {\bf p})$ variables are used, this 
equation is a seven-dimensional differential equation (actually  an integro-differential equation 
because of Eq. (\ref{dvint})) containing partial derivatives with respect to time and to the six 
variables $r_i$ and $p_i$.
        The time derivative can be eliminated by using the Fourier transform (\ref{fft}), while 
the properties of Poisson brackets suggest that  some simplification might be achieved simply  by 
making a change of variables to generalized coordinates and momenta. The new coordinates should be 
chosen in such a way to include the maximum number  of constants of motion of the unperturbed 
Hamiltonian $h_0$ so that the corresponding variable will not contribute to the Poisson bracket. 
For motion in a central force field a convenient set of generalized coordinates is ($\epsilon,
\lambda,r,\alpha,\beta,\gamma$), where $\epsilon$ is the particle energy (\ref{qpe}), $\lambda$ 
the magnitude of its angular momentum, $r$ the radial coordinate, and $\alpha,\beta,\gamma$ the Euler angles associated with the rotation of the frame of 
Cartesian coordinates necessary to align the $z$-axis of the lab frame to the particle angular 
momentum
$\vec{\lambda}$ and the $y$-axis of the lab frame with the ${\bf r}$ vector specifying the 
instantaneous position of a particle with respect to the force centre (see Fig. 1). 

\begin{figure}[t]
\label{fig1}
\begin{center}
\includegraphics[width=3in]{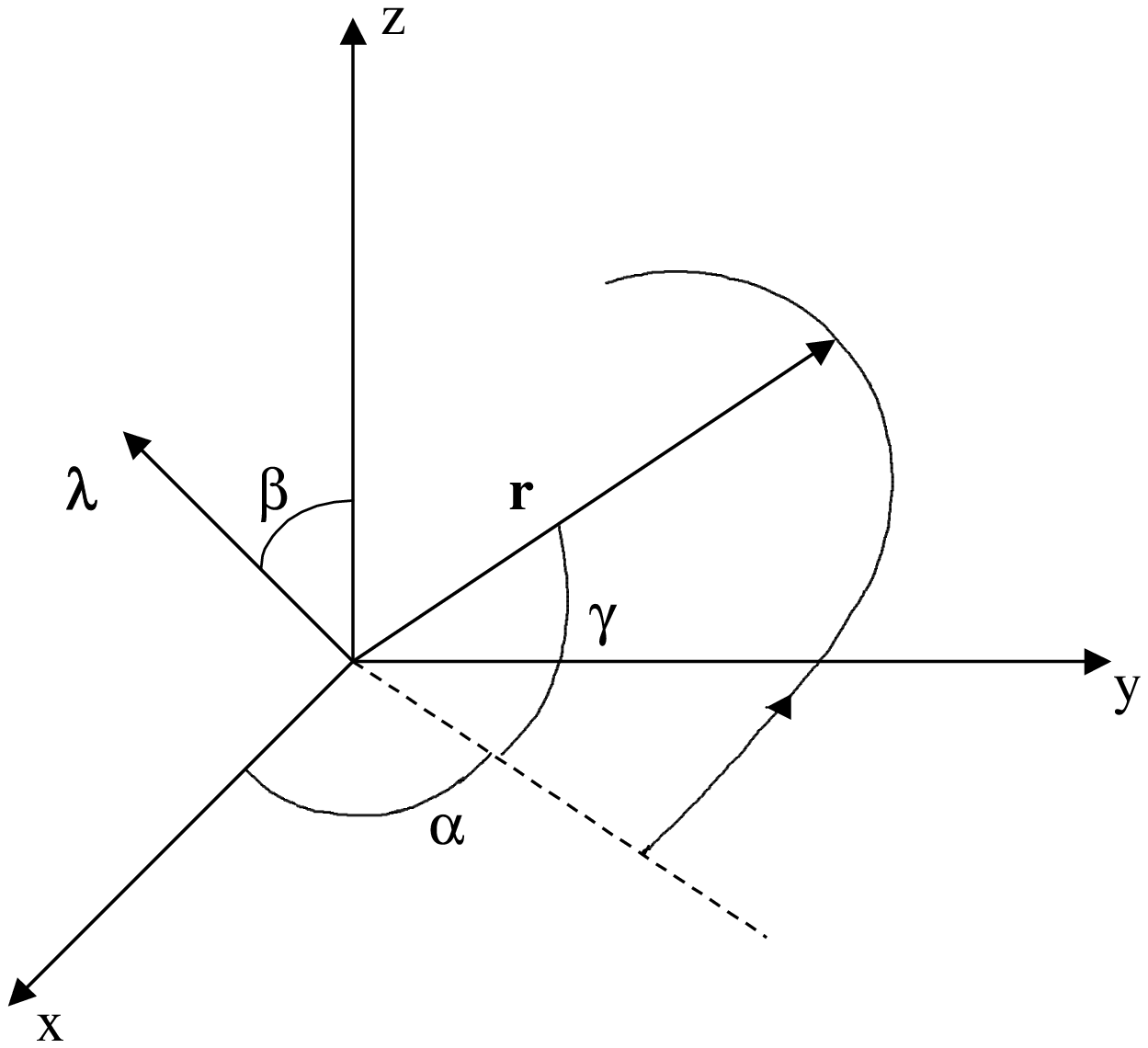}
\caption{Angular elements of the orbit. For a particle moving in a central force field, the angles $\alpha$ and $\beta$ are constant.}
\end{center}
\end{figure}

Since four of 
the six new coordinates are contants of the motion,  Eq. (\ref{lve2}) simplifies considerably and 
becomes
\beq
\label{boh}
\frac{\partial\,\delta f}{\partial t}+\frac{\partial\,\,\delta f}{\partial r}\dot r+\frac{\partial\,
 \delta f}{\partial \gamma}\dot \gamma=
-F'(\epsilon)[\frac{\partial\,\delta h}{\partial r}\dot r+\frac{\partial\, \delta h}
{\partial \gamma}\dot \gamma]\,.
\eeq
By using
\bea
&&\dot \gamma=\frac{\lambda}{mr^2}\,, \\
&&\dot r=\pm v_r(\epsilon,\lambda,r)\,,
\eea
with
\beq
\label{vrad}
v_r(\epsilon,\lambda,r)=\sqrt{\frac{2}{m}[\epsilon-V_0(r)-\frac{\lambda^2}{2mr^2}]}
\eeq
the magnitude of the radial velocity, we obtain
\beq
\label{int}
-i\omega\delta f\pm v_r(\epsilon,\lambda,r)\frac{\partial\,\delta f}{\partial r}+\frac{\lambda}
{mr^2}\frac{\partial \,\delta f}{\partial \gamma}=
F'(\epsilon)[\pm v_r(\epsilon,\lambda,r) \frac{\partial\,\delta h}{\partial r}+\frac{\lambda}
{mr^2}\frac{\partial\, \delta h}{\partial \gamma}]\,
\eeq
for the Fourier-transformed linearized Vlasov equation.

This equation still contains partial derivatives with respect to the two time-dependent variables 
$r$ and $\gamma$, but the derivative with respect to $\gamma$ can be eliminated by means of an 
appropriate partial-wave expansion. The usual partial-wave expansion
\beq
\label{pwe1}
\delta f({\bf r},{\bf p},\omega)=\sum_{LM}\delta f_{LM}(r,{\bf p},\omega) Y_{LM}(\vartheta,
\varphi)\,,
\eeq
where $(\vartheta,\varphi)$ are the polar angles of the vector ${\bf r}$, is a first step in this direction, but it does not solve the 
problem since the coefficients $\delta f_{LM}(r,{\bf p},\omega)$ still depend on the direction 
of the vector ${\bf p}$, however the following well known transformation property of the spherical 
harmonics $Y_{LM}(\vartheta,\varphi)$ under the rotation specified by the Euler angles $\alpha,
\beta,\gamma$ can help us (see e.g. (\cite{bsa}, p. 28) \footnote{In  this rather technical aspect 
the derivation of \cite{bri} differs by that of \cite{pol}, which is less straightforward (cf. 
also \cite{ber2}).}:
\beq
\label{tran}
Y_{LM}(\vartheta,\varphi)=\sum_{N=-L}^{L}{\Big (}{\cal D}^L_{MN}(\alpha\beta,
\gamma){\Big )}^*Y_{LN}(\vartheta',\varphi')\,.
\eeq

In the new reference frame the particle is on the $y$ axis,  so $\vartheta'=\frac{\pi}{2}$ and 
$\varphi'=\frac{\pi}{2}$ and the only time-dependent angle on the righ-hand-side of 
Eq. (\ref{tran}) is $\gamma$. The functions
${\cal D}^L_{MN}(\alpha\beta,\gamma)$ are the coefficients of the rotation matrices and their 
explicit $\gamma$ dependence, of the kind $e^{-iN\gamma}$, can be exploited to eliminate the 
$\gamma$-derivative in Eq. (\ref{int}) , giving the following one-dimensional equations
\beq
\label{fin}
\frac{\partial}{\partial r}\delta f^{L\pm}_{MN}\mp A_N\delta f ^{L\pm}_{MN}=B^{L\pm}_{MN}\,,
\eeq
with
\beq
\label {afu}
A_N(\epsilon,\lambda,r,\omega)=\frac{i\omega}{v_r(\epsilon,\lambda,r)}-\frac{iN}{v_r(\epsilon,
\lambda,r)}
\frac{\lambda}{mr^2}
\eeq
and
\beq
\label{bfu}
B^{L\pm}_{MN}(\epsilon,\lambda,r,\omega)=F'(\epsilon)[\frac{\partial}{\partial r}\pm \frac{iN}
{v_r(\epsilon,\lambda,r)}\frac{\lambda}{mr^2}][\beta Q_{LM}(r)+
\delta V^{int}_{LM}(r,\omega)]\,,
\eeq
for the coefficients $\delta f^{L\pm}_{MN}(\epsilon,\lambda,r,\omega)$ 
of the expansion \footnote{ Note that, contrary to what is done in Ref. 
\cite{bri}, here  the factor $Y_{LN} (\frac{\pi}{2},\frac{\pi}{2})$ is not 
included in our definition of the coefficients $\delta f^{L\pm}_{MN}$.}
\bea
\label{exp}
\delta f({\bf r},{\bf p},\omega)=&&\sum_{LMN}[\delta f^{L+}_{MN}(\epsilon,\lambda,r,\omega) \theta(p_r)+
\delta f^{L-}_{MN}(\epsilon,\lambda,r,\omega) \theta(-p_r)]\nonumber\\
&&{\Big (}{\cal D}^L_{MN}(\alpha\beta,\gamma){\Big )}^*
Y_{LN}(\frac{\pi}{2},\frac{\pi}{2})\,.
\eea
In Eq. (\ref{bfu}) the functions $Q_{LM}(r)$ and $\delta V^{int}_{LM}(r,\omega)$  are the 
coefficients of a multipole expansion similar to (\ref{pwe1}) for the external driving field and 
for the induced mean-field fluctuation, respectively. In Eq. (\ref{exp}) instead, the $\theta$-functions
are the usual step function $\theta(x)=1$for $x>0$ and $\theta(x)=0$ otherwise, while $p_r$ is the radial component of the particle momentum: $p_r=\pm mv_r$.

        Thus,  by making an appropriate change of variables (and by taking the time Fourier 
transform), the initial seven-dimensional differential equation (\ref{lve2}) has been reduced to 
the system of two (coupled) one-dimensional differential equations (\ref{fin}). These two 
equations, involving the distributions $\delta f^+$ and $\delta f^-$  of particles with both signs 
of the radial velocity, are coupled by the mean-field fluctuation, as shown explicitly by the 
following expression:
\bea
\label{delv}
\delta V^{int}_{LM}(r,\omega)=&&\frac{8\pi^2}{2L+1}\sum_{N=-L}^{L}|Y_{LN}(\frac{\pi}{2},
\frac{\pi}{2})|^2
\int d\epsilon\int d\lambda \lambda\int \frac{dr'}{v_r(\epsilon,\lambda,r')} v_L(r,r')\nonumber\\
&&[\delta f^{L+}_{MN}(\epsilon,\lambda,r',\omega)+\delta f^{L-}_{MN}(\epsilon,\lambda,r',
\omega)]\,.
\eea
For most effective interactions $v_L(r,r')$, because of this coupling the  solution of 
Eq.(\ref{fin}) can only be given  in implicit form, however, an explicit solution can be obtained  
if  we neglect the term $\delta V^{int}_{LM}(r,\omega)$ in Eq.(\ref{bfu}). Following Ref. 
\cite{bri}, we refer to this  as the zero-order approximation and recall here the solution details.

In order to solve Eq. (\ref{fin}), we must first specify the 
boundary conditions satisfied by $\delta f^+(r)$ and $\delta f^-(r)$ at the turning points 
$r_1$ and $r_2$. The boundary conditions used in \cite{bri} were:
\bea
\label{bc1}
&&\delta f^{L+}_{MN}(\epsilon,\lambda,r_1,\omega)=\delta f^{L-}_{MN}(\epsilon,\lambda,r_1,
\omega)\,,\\
\label{bc2}
&&\delta f^{L+}_{MN}(\epsilon,\lambda,r_2,\omega)=\delta f^{L-}_{MN}(\epsilon,\lambda,r_2,\omega)
\,,
\eea
their physical meaning is  that, at the turning points the radial motion of the particles simply 
reverses. For the square-well potential, the condition at the outer turning point implies a 
mirror reflection of the nucleons on the static equilibrium nuclear surface. As we shall see, 
in the sharp-surface case there are reasons for modifying this boundary condition, however for 
the moment we assume a diffused surface and determine the solution of Eq.(\ref{fin}) by using 
the boundary conditions (\ref{bc1}) and (\ref{bc2}). The solution can be written as (in slightly 
simplified  notation)
\beq
\label{sol1}
\delta f^{L\pm}_{MN}(\epsilon,\lambda,r,\omega)=e^{\pm i\phi_{N}(r,\omega)}[
\int_{r_{1}}^{r}d r'{ B}^{L\pm}_{MN}(r')
e^{\mp i\phi_{N}(r',\omega)]}\:+C_{\pm}(\epsilon,\lambda,\omega)]\,,
\eeq

with
\beq
\phi_{N}(r,\omega)=-i\int_{r_{1}}^{r}dyA_{N}(y)=
\omega\tau(r)-N\gamma(r)\,,
\eeq
\bea
\label{tau}
&&\tau(r)=\int_{r_1}^{r} dr'\frac{1}{v_r(\epsilon,\lambda,r')}\,,\\
\label{gam}
&&\gamma(r)=\int_{r_1}^{r} dr'\frac{\lambda}{mr'^2}\frac{1}{v_r(\epsilon,\lambda,r')}\,.
\eea

The functions $C_{\pm}(\epsilon,\lambda,\omega)$ play the role of integration contants and are 
determined by the
boundary conditions. The inner boundary  condition (\ref{bc1}) implies
\beq
\label{c1}
C_{-}=C_{+}\,,
\eeq

while the outer boundary condition (\ref{bc2}) implies
\bea
\label{c2}
&&e^{i\phi_{N}(r_2,\omega)}[\int_{r_{1}}^{r_2}dr'{ B}^{L+}_{MN}(r')e^{-i\phi_{N}(r',\omega)}\:+\:C_{+}]=
\nonumber\\
&&e^{-i\phi_{N}(r_2,\omega)}[\int_{r_{1}}^{r_2}dr'{ B}^{L-}_{MN}(r')e^{i\phi_{N}(r',\omega)}\:+\:C_{+}]\,.
\eea
 Defining
\beq
D_{\pm}(\epsilon,\lambda,\omega)=\int_{r_{1}}^{r_2}dr'{ B}^{L\pm}_{MN}(r')
e^{-i\phi_{N}(r',\omega)}\,,
\eeq
gives
\beq
e^{2i\phi_{N}(r_2,\omega)}[D_{+}\,+\,C_{+}]=D_{-}\,+C_{+}\,,
\eeq
that is
\beq
\label{const}
C_{+}(\epsilon,\lambda,\omega)=
\frac{e^{2i\phi_{N}(r_2,\omega)}D_{+}-D_{-}}{1-e^{2i\phi_{N}(r_2,\omega)}}=
C_{-}(\epsilon,\lambda,\omega)\,.
\eeq

        The most interesting property of the solution (\ref{sol1}) is its pole structure in the 
complex-$\omega$ plane, which is entirely determined by the poles of the functions 
$C_{\pm}(\epsilon,\lambda,\omega)$, that is, by the vanishing of the denominator in 
Eq. (\ref{const}).
        This happens whenever $2\phi_N(r_2,\omega)=n\,2\pi$, with integer $n$.
        This is the  point where the finite size of the system plays a crucial role since the 
eigenfrequencies of the density fluctuations in the zero-order approximation  are determined  
by  the condition
        \beq
        \omega [2\tau(r_2)]-N[2\gamma(r_2)]=n\,2\pi\,.
        \eeq
and the period of radial motion $T(\epsilon,\lambda)=2\tau(r_2)$ depends on the size of the system.
      For motion in a central potential, these eigenfrequencies are characterized by the two 
integers $n$ and $N$:
\beq
\label{eif}
\omega_{nN}(\epsilon,\lambda)=n\omega_0 +N\omega_\varphi\,,
\eeq
with
\bea
\label{orad}
&&\omega_0(\epsilon,\lambda)=\frac{2\pi}{T(\epsilon,\lambda)}\,,\\
\label{oang}
&&\omega_\varphi(\epsilon,\lambda)=\frac{\gamma(r_2)}{\tau(r_2)}\,.
\eea
 the frequencies of radial and angular motion for a particle with energy $\epsilon$ and angular 
momentum magnitude  $\lambda$.
The eigenfrequencies (\ref{eif}) can be related to the {\em difference} of single-particle energy 
levels $\epsilon_{nl}$ in a central potential since, for large quantum numbers 
(\cite{bm2}, p. 579),
\beq
\label{ediff}
\epsilon_{n'l'}-\epsilon_{n"l"}\approx(n'-n")\frac{\partial \epsilon}{\partial n}+(l'-l")
\frac{\partial \epsilon}{\partial l}=\hbar \omega_{n'-n"\,l'-l"}\,.
\eeq

Thus the integer $n$ can be interpreted as the difference between radial quantum numbers in a 
single-particle excitation and the integer $N$ as the difference between the corresponding orbital 
quantum numbers.

The time-dependent density fluctuations can be obtained from the functions 
$\delta f^{L\pm}_{MN}(\epsilon,\lambda,r,\omega)$ by contour integration in the 
complex-$\omega$ plane. For the zero-order fluctuations we obtain:
\bea
\delta f^{0L\pm}_{MN}(\epsilon,\lambda,r,t)&&=0\qquad\qquad\qquad\qquad\qquad\qquad\qquad
\quad{\rm for}\quad t<0\,,\\
\label{sec}
&&=\frac{1}{2\pi}\int_{-\infty}^{+\infty} d\omega e^{-i\omega t}\delta f^{0L\pm}_{MN}(\epsilon,
\lambda,r,\omega)\quad{\rm for}\quad t>0\,,
\eea
with
\beq
\label{zos}
\delta f^{0L\pm}_{MN}(\epsilon,\lambda,r,\omega)=-\beta F'(\epsilon)\sum_{n=-\infty}^{\infty}
\omega_{nN}
e^{\pm i\phi_{nN}(r)}\frac{Q_{nN}(\epsilon,\lambda)}{\omega-\omega_{nN}+i\varepsilon}\,,
\eeq
 \beq
 \phi_{nN}(r)=\omega_{nN}\tau(r)-N\gamma(r)
 \eeq
 and
 \beq
 \label{fco}
 Q_{nN}(\epsilon,\lambda)=\frac{1}{\tau(r_2)}\int_{r_1}^{r_2} dr\frac{Q_{LM}(r)}{v_r(\epsilon,
\lambda,r)}\cos[\phi_{nN}(r)]\,.
 \eeq

This result is  obtained from Eq. (\ref{sol1}) by closing the integration path in the lower part 
of the complex-$\omega$ plane in the integral (\ref{sec}).

 The coefficients (\ref{fco}) correspond to the classical limit of the quantum matrix elements of 
the external field
 (\ref{vext}) \cite{mig}.

        Thus we have seen that, for a spherical nucleus, the linearized Vlasov equation can be 
solved explicitly in the approximation in which the mean-field fluctuation is neglected. This 
zero-order solution can be used as a starting point for solving also the more general problem in 
which the mean field fluctuation $\delta V^{int}$ is taken into account. The zero-order solution 
is most conveniently expressed in terms of a semiclassical propagator ( obtained from Eq. 
(\ref{zos}), see \cite{bri}) which is analogous to the quantum particle-hole propagator
\bea
\label{zop}
D^0_L(r,r',\omega)=&&\frac{8\pi^2}{2L+1}\sum_{n=-\infty}^{+\infty}\sum_{N=-L}^L|Y_{LN}
(\frac{\pi}{2},\frac{\pi}{2})|^2\int d\epsilon F'(\epsilon)\int d\lambda \lambda\nonumber\\
&&\frac{1}{T(\epsilon,\lambda)}\,\frac{\cos[\phi_{nN}(r)]}{r^2 v_r(\epsilon,\lambda,r)}\,
\frac{\omega_{nN}}{\omega_{nN}-(\omega+i\varepsilon)}\,
\frac{\cos[\phi_{nN}(r')]}{r'^2 v_r(\epsilon,\lambda,r')}\,.
\eea

Taking into account also the mean-field fluctuation (\ref{dvint})       gives the collective 
response of the system and leads in general to an integral equation  which is analogous to the 
RPA equation for the quantum propagators
\beq
\label{rpa}
D_L(r,r',\omega)=D^0_L(r,r',\omega)+\int dx x^2 \int dyy^2D^0_L(r,x,\omega) v_L(x,y)D_L(y, r',
\omega)\,.
\eeq
The only differences between the present expression and the quantum result are that here the 
propagator $D^0$ is given in semiclassical approximation and, of course, the exchange (Fock) 
term is missing. The integral equation (\ref{rpa}) can be easily solved numerically 
(the zero-order propagator is actually simpler than suggested by Eq. (\ref{zop}) since, for small values of $\omega$, the 
infinite sum over $n$ can be approximated with a sum over a few terms around $n=0$ with very 
good accuracy). The collective (multipole) response function is then given by
\beq
\label{coll}
{\cal R}_L(\omega)=\int dr r^2\int dr' r'^2 Q_{LM}(r)D_L(r,r',\omega)Q_{LM}(r')\,,
\eeq
with $D_L$ solution of (\ref{rpa}), and the corresponding strength function by
\beq
\label{stf}
S_L({\hbar\omega})=-\frac{1}{\pi}{\rm Im}{\cal R}_L(\hbar\omega)
\eeq
(for a spherical system the response is independent of $M$).

However there is a special case in which also the collective solution of the linearized Vlasov equation can be obtained explicitly. This happens if the interaction between particles is supposed 
to be of the separable multipole-multipole type:
\beq
\label{sepi}
v({\bf r},{\bf r'})=\sum_{LM}\kappa_L r^Lr'^LY_{LM}({\bf \hat r})Y^*_{LM}({\bf \hat r}')\,
\eeq
and the external field is also of the multipole type (\ref{mul}). In this case Eqs. (\ref{rpa}) 
and (\ref{coll}) give immediately
\beq
\label{sepre}
{\cal R}_L(\omega)=\frac{{\cal R}^0_L(\omega)}{1-\kappa_L{\cal R}^0_L(\omega)}\,,
\eeq
with the zero-order response function ${\cal R}^0_L(\omega)$ given by
\bea
\label{zor}
{\cal R}^0_L(\omega)=&&\frac{1}{\beta}\frac{8\pi^2}{2L+1}\sum_{N=-L}^L|Y_{LN}(\frac{\pi}{2},
\frac{\pi}{2})|^2\int d\epsilon \int d\lambda \lambda \nonumber\\
&&\int_{r_1}^{r_2} dr\frac{Q_{LM}(r)}{v_r(\epsilon,\lambda,r)}[\delta f^{0L+}_{MN}(\epsilon,
\lambda,r,\omega)+
\delta f^{0L-}_{MN}(\epsilon,\lambda,r,\omega)]\,.
\eea
 Moreover the  fluctuation $\delta f$ can also be obtained in explicit form, like $\delta f^0$.  
For the separable interaction (\ref{sepi}), Eq. (\ref{delv}) for the mean-field fluctuation gives
\beq
\label{delvsep}
\delta V^{int}_{LM}(r,\omega)=\beta\kappa_Lr^L{\cal R}_L(\omega)\,,
\eeq
while Eq. (\ref{bfu}) gives
\beq
\label{bfusep}
B^{L\pm}_{MN}(\epsilon,\lambda,r,\omega)=F'(\epsilon)[\frac{\partial}{\partial r}\pm 
\frac{iN}{v_r(\epsilon,\lambda,r)}\frac{\lambda}{mr^2}][\beta r^L+\beta\kappa_L
r^L{\cal R}_L(\omega)]\,,
\eeq
 and, from Eq.(\ref{fin}), we get $\delta f^{L\pm}_{MN}(\epsilon,\lambda,r,\omega)=
\delta f^{0L\pm}_{MN}(\epsilon,\lambda,r,\omega)[1+\kappa_L{\cal R}_L(\omega)]$ or, 
by using Eq. (\ref{sepre}),
\beq
\label{sepdf}
\delta f^{L\pm}_{MN}(\epsilon,\lambda,r,\omega)=\frac{\delta f^{0L\pm}_{MN}(\epsilon,\lambda,r,
\omega)}{1-\kappa_L{\cal R}^0_L(\omega)}\,.
\eeq

\subsection{Action-angle variables}

Up to now we have assumed a spherically symmetric equilibrium mean field, however the method 
outlined here for the solution of the linearized Vlasov equation is valid also for a wider class 
of physical systems. This method, which is based on the use of generalized coordinate and momenta 
in order to simplify the Vlasov equation, can actually be used for all systems which are described 
by an integrable equilibrium Hamiltonian $h_0({\bf r},{\bf p})$. Such a Hamiltonian includes also some deformed systems \cite{dmb}.
For integrable systems, it is convenient to introduce action-angle variables 
$({\bf I},{\bf \Phi})$ instead of $({\bf r},{\bf p})$ since in this case the action variables  
$I_\alpha$ are constants of the motion, while the angle variables $\Phi_{\alpha}$  are linear 
functions of time (see for example Ref. \cite{gol}, p. 457).
An important property of these variables is that the motion is periodic
in the angle variables with period $2\pi$. Consequently the field felt
by a particle that is moving along  a trajectory determined by an integrable Hamiltonian can be Fourier expanded as
\beq
\label{fou1}
V^{ext}({\bf r},\omega)=\beta\sum_{{\bf n}}Q_{\bf n}({\bf
I}) e^{i{\bf n}\cdot {\bf \Phi}}\,,
\eeq
and
\beq
\label{fou2}
\delta V^{int}({\bf r},\omega)=\sum_{{\bf n}}\delta V_{\bf n}({\bf
I},\omega) e^{i{\bf n}\cdot {\bf \Phi}}
\eeq
where ${\bf n}$ is a three-dimensional vector with integer components.

The phase-space-density fluctuation can also be expanded in the same way:
\beq
\label{fou3}
\delta f({\bf r},{\bf p},\omega)=\sum_{\bf n}\delta f_{\bf n}({\bf I},\omega)
e^{i{\bf n}\cdot {\bf \Phi}}\,
\eeq

and the linearized Vlasov equation gives the following equation for the coefficients 
$\delta f_{\bf n}({\bf I},\omega)$:
\beq
\label{sol}
\delta f_{\bf n}({\bf I},\omega)=F'(\epsilon)
{\Big [}\beta Q_{\bf n}({\bf I})
+\delta V^{int}_{\bf n}({\bf I},\omega){\Big ]}\frac{{\bf n}\cdot
\vec{\omega}} {{\bf n}\cdot \vec{\omega}-(\omega +i\varepsilon)}\,,
\eeq
where the vector $\vec{\omega}$ has components
\beq
\omega_\alpha= \frac{\partial h_{0}({\bf I})}{\partial I_{\alpha}}\,.
\eeq
Again, Eq. (\ref{sol}) gives only an implicit solution of the Vlasov equation since 
the mean-field fluctuation $\delta V^{int}$ depends on $\delta f$. If the term $\delta V^{int}$ 
is neglected, then Eq. (\ref{sol}) is an explicit solution that, in the case of spherical 
systems agrees with the zero-order solution (\ref{zos}). Generally speaking, the vector 
$\vec{\omega}$ has three components,but in spherical systems there are only two basic 
eigenfrequencies: $\omega_0$ and $\omega_\varphi$. This is because spherical systems are 
over-integrable and this implies that one of the angle variables is also a constant of the 
motion. The coefficients $Q_{\bf n}({\bf I})$ in Eq. (\ref{sol}) are given by
\beq
\label{fou4}
Q_{\bf n}({\bf I})=\frac{1}{2\pi^3}\int d{\bf \Phi} e^{-i{\bf n}\cdot{\bf \Phi}} Q({\bf r})
\eeq
and correspond to the quantum matrix elements of the operator $Q({\bf r})$.

\subsection{Sharp and moving surface}

        With the aim of establishing a link between the present microscopic 
theory of nuclear excitations and the macroscopic description given by the 
liquid-drop model (see e. g. \cite{bm2}, Appendix 6A) the authors of \cite{ads}
studied in detail the model in which the equilibrium mean 
field is approximated by a square-well potential: $V_0(r)=-V_0\vartheta(R-r)$.
They noticed  that in this case the boundary condition (\ref{bc2}) corresponds
to a mirror reflection of nucleons when they reach the static nuclear surface
at $r=R$. With that boundary condition the sharp-surface model allows only for 
compressional excitations, while a liquid drop has both surface and 
compression modes.  They argued that, in order to allow also for a microcopic 
description of surface modes, in the sharp-surface case, the boundary condition (\ref{bc2}) 
should be modified and they proposed to replace it with the following 
(moving-surface) boundary condition:
\beq
\label{bc3}
\delta \tilde f^{L+}_{MN}(\epsilon,\lambda,R,\omega)=
\delta \tilde f^{L-}_{MN}(\epsilon,\lambda,R,\omega)+
F'(\epsilon)2m v_r(\epsilon,\lambda,R) i\omega  \delta R_{LM}(\omega)\,.
\eeq
Whenever the same symbol is used,
we  put a tilde over the quantities 
evaluated with the moving-surface boundary condition (\ref{bc3}), to 
distingush them from the corresponding quantities satisfying the fixed-surface 
boundary condition (\ref{bc2}).
The physical picture behind  this new boundary condition still corresponds to 
a mirror reflection of nucleons at the nuclear surface $r=R$, but in a 
reference frame that is moving in the radial direction at a speed 
$\dot{R}(\vartheta,\varphi,t)$. The radial momentum transfered by the moving 
surface to the impinging nucleon will differ from that occurring in a 
collision with the static surface and this modifies the momentum distribution 
of nucleons according to (\ref{bc3}),  in  first approximation.
Formally, in this approach the nuclear surface is allowed to vibrate 
(for $t>0$) according to the usual liquid-drop model expression:
\beq
\label{ldmr}
R(\vartheta,\varphi,t)=R+\sum_{LM}\delta R_{LM}(t) Y_{LM}(\vartheta,\varphi)\,,
\eeq
giving
\beq
\label{ldmv}
\dot{R}(\vartheta,\varphi,t)=\sum_{LM}\delta \dot{R}_{LM}(t) Y_{LM}(\vartheta,
\varphi)
\eeq
for the surface speed and
\beq
\label{ftsv}
\dot{R}(\vartheta,\varphi,\omega)=\sum_{LM}-i\omega\delta {R}_{LM}(\omega) 
Y_{LM}(\vartheta,\varphi)
\eeq
for its time Fourier transform, thus leading to the extra term in Eq. 
(\ref{bc3}).

        The new collective coordinates $\delta {R}_{LM}(\omega)$ are still to 
be determined. In \cite{ads} this has been done by recalling that in the 
liquid-drop model a change in the curvature radius 
of the surface results in a change of pressure given by (see Eq. (6A-57) 
of \cite{bm2})
\beq
\label{delp}
\delta {\cal P}(R,\theta,\varphi,\omega)=\sum_{LM}C_{L}\frac{\delta
R_{LM}(\omega)}{R^{4}} Y_{LM}(\theta,\varphi)\,,
\eeq
with the restoring force parameter $C_L$ that can be related to the 
phenomenologically determined surface tension parameter 
$\sigma\approx 1\:{\rm MeV\,fm^{-2}}$. If the Coulomb repulsion between 
protons is neglected, this relation is simply
\beq
\label{rfp}
C_L=\sigma R^2 (L-1)(L+2)\,,
\eeq
while taking into account also the Coulomb interaction gives an additional 
contribution to $C_L$ (see \cite{bm2}, p.660).  The pressure fluctuation 
(\ref{delp}) can also be related to the appropriate component of the pressure 
tensor (generalized to Fermi liquids, see \cite{lif}, Sect. 74)
\beq
\label{prr}
\delta{\cal P}(R,\theta,\varphi,\omega)=\int d{\bf p} m
v_{r}^2{\big (} \delta \tilde  f({\bf R},{\bf p},\omega)-F'(\epsilon)\delta
\tilde V^{int}({\bf R},\omega){\big )}\,.
\eeq
By equating the pressure fluctuations given by Eq. (\ref{delp}) and 
(\ref{prr}), the collective coordinates $\delta {R}_{LM}(\omega)$ can be 
related to the phase-space-density fluctuation:
\bea
\label{delrlm}
&&\delta R_{LM}(\omega)=
\frac{8\pi^{2}}{2L+1}\frac{R^{2}}{C_{L}}\sum_{N=-L}^{L}
{\Big |}Y_{LN}(\frac{\pi}{2},\frac{\pi}{2}){\Big |}^{2}
\int d\epsilon \int d\lambda \lambda m v_{r}(\epsilon,\lambda,R)\nonumber\\
&&{\Big [}\delta \tilde f^{L+}_{MN}(\epsilon,\lambda,R,\omega)+\delta 
\tilde f^{L-}_{MN}(\epsilon,\lambda,R,\omega)
-2F'(\epsilon)\delta \tilde V^{int}_{LM}(R,\omega){\Big ]}\,.
\eea
The internal part of the mean-field fluctuation $\delta \tilde V^{int}_{LM}(r,\omega)$ in the moving-surface case will be specified better in  Appendix A. This can be done most easily by assuming a separable interaction of the kind (\ref{sepi}). We have already seen that such an effective interaction leads to simple analytical expressions for the multipole response function and for the solution of the linearized Vlasov equation with fixed-surface boundary conditions. The same happens also in the moving-surface case, although the final expressions are somewhat more involved. Since the explicit derivation of the moving-surface multipole response function is rather lengthy, we report here only the final result
\bea
\label{msresp}
&&{\tilde{\cal R}}_L(\omega)={\cal R}_L(\omega)\\
&&-\frac{1}{1-\kappa_{L} {\cal R}_{L}^{0}(\omega)}\: 
\frac{[\chi^{0}_{L}(\omega)+\frac{3}{4\pi}A\kappa_{L}R^{L}
{\cal R}_{L}^{0}(\omega)]^{2}}{[C_{L}-\chi_{L}(\omega)][1-\kappa_{L} 
{\cal R}_{L}^{0}(\omega)]+\kappa_{L}[\chi^{0}_{L}(\omega)+
\frac{3}{4\pi}AR^{L}]^{2}}\,,\nonumber
\eea
and  outline it in  Appendix A.
The response function ${\cal R}_L(\omega)$ in the equation above is still that given by Eq. (\ref{sepre}), while  the functions $\chi_{L}(\omega)$ and $\chi^{0}_{L}(\omega)$ are defined as\footnote { In Refs. \cite{adm2,adm5}, a different normalization of $\chi^0_L$ has been used.}
 \bea
&&\label{chil}
\chi_{L}(\omega)=\frac{8\pi^{2}}{2L+1}{R^{2}}\sum_{N=-L}^{L}
{\Big |}Y_{LN}(\frac{\pi}{2},\frac{\pi}{2}){\Big |}^{2}\nonumber\\
&&\int d\epsilon \int d\lambda \lambda 
2F'(\epsilon)\cot[\phi_N(R,\omega)][mv_r(\epsilon,\lambda,R)]^2 \omega.
\eea
and
\bea
\label{chi0l}
&&\chi^{0}_{L}(\omega)=
\frac{1}{\beta}\frac{8\pi^{2}}{2L+1} R \sum_{N=-L}^{L}
{\Big |}Y_{LN}(\frac{\pi}{2},\frac{\pi}{2}){\Big |}^{2}\nonumber\\
&&\int d\epsilon \int d\lambda \lambda m v_{r}(\epsilon,\lambda,R)
{\Big [}\delta f^{0L+}_{MN}(\epsilon,\lambda,R,\omega)+\delta 
f^{0L-}_{MN}(\epsilon,\lambda,R,\omega){\Big ]}\,.
\eea

\section{Isoscalar excitations in heavy nuclei}

	In the remaining part of this paper we review some applications of the theory outlined in the first part to the study of isoscalar nuclear response . Our approximation for the mean field ( a square-well potential) is not particularly realistic since it neglects the surface diffusion and the assumed residual interaction
of the multipole-multipole type is also rather special, however these approximations have the advantage of leading to simple analytical formulae for the nuclear response functions. Our semiclassical approach shows that some feature of the nuclear response that are usually ascribed to quantum effects can be understood in terms of classical concepts like nucleon trajectories. The correspondence between  shell effects (not included in our treatment) and the properties of classical trajectories helps to shed a new light on  the nuclear response at low energy.

	Our starting point is the zero-order response function (\ref{zor}). By using the explicit expression
(\ref{zos}) of the phase-space-density fluctuations, with $F'(\epsilon)=-\frac{4}{(2\pi)^3}\delta(\epsilon_F-\epsilon)$, this equation gives	
\beq
\label{zor2}
{\cal R}^0_L(s)=\frac{9A}{8\pi}\frac{R^{2L}}{\epsilon_F}\sum_{N=-L}^LC_{LN}^2 \sum_{n=-\infty}^{+\infty}
\int_0^1 dx x^2  \frac{s_{nN}(x)}{s-s_{nN}(x)+i\varepsilon}(Q_{nN}^{(L)}(x)/R^L)^2\,,
\eeq	
with  $x=\sin\alpha$, $\cos\alpha=\lambda/(p_FR)$, $s=\omega/(v_F/R)$, $v_F=p_F/m$,
\beq
s_{nN}(x)=\frac{n\pi+N\alpha(x)}{x}\,,
\eeq	
\beq
\label {fc}
Q_{nN}^{(k)}(x)=\frac{2}{T}\int_{r_1}^{R} d r \frac{r^k}{v_r(\epsilon_F,\lambda,r)}\cos[\phi_{nN}(r)]\,,
\eeq	
and
\beq
\label{sh}
C_{LN}^2=\frac{4\pi}{2L+1}|Y_{LN}(\frac{\pi}{2},\frac{\pi}{2})|^2\,.
\eeq
Instead of the frequency $\omega$, we have introduced the dimensionless parameter $s$ and instead of the particle angular momentum $\lambda$, the parameter $x=\sin\alpha$. 

The Fourier coefficients (\ref{fc}) can be easily evaluated explicitly, the expressions needed here are grouped together in Appendix B.

In terms of the new dimensionless variables, the auxiliary  functions  $\chi^0_L$ and $\chi_L$ appearing in eq. (\ref{msresp}) read
\beq
\label{chizs}
\chi^0_L(s)=\frac{9A}{4\pi}R^L\sum_{nN}C^2_{LN}\int_0^1dxx^2 s_{nN}(x)\frac{(-)^n Q^{(L)}_{nN}(x)/R^L}{s+i\varepsilon-s_{nN}(x)}
\eeq

and
\beq
\chi_L(s)=-\frac{9A}{2\pi}\epsilon_F s\sum_{nN}C^2_{LN}\int_0^1dxx^2 \frac{1}{s+i\varepsilon-s_{nN}(x)}\,.
\eeq
In the last equation we have used the pole expansion of the cotangent: $$\cot z=\sum_{n=-\infty}^{+\infty}
\frac{1}{z-n\pi}\,.$$

\subsection{Monopole response}

This channel corresponds to a compression mode. For these modes the radial dependence of the external field is not given by Eq. (\ref{mul}), but rather by
\beq
Q({\bf r})=r^{L+2} Y_{LM}({\bf \hat r}).
\eeq

As a consequence the zero-order response function (\ref{zor2}) willl involve the Fourier coefficients $Q_{nN}^{(L+2)}(x)$ instead of $Q_{nN}^{(L)}(x)$ and the same is true for the auxiliary function $\chi^0_L(s)$ in Eq. (\ref{chizs}). Accordingly, the moving-surface response function  (\ref{msresp}) is also slightly changed  to
\bea
\label{cmmsresp}
&&{\tilde{\cal R}}_L(\omega)={\cal R}_L(\omega)\\
&&-\frac{1}{1-\kappa_{L} {\cal R}_{L}^{0}(\omega)}\: 
\frac{[\chi^{0}_{L}(\omega)+\frac{3}{4\pi}A\kappa_{L}R^{L+2}
{\cal R}_{L}^{0}(\omega)]^{2}}{[C_{L}-\chi_{L}(\omega)][1-\kappa_{L} 
{\cal R}_{L}^{0}(\omega)]+\kappa_{L}[\chi^{0}_{L}(\omega)+
\frac{3}{4\pi}AR^{L+2}]^{2}}\,.\nonumber
\eea

\begin{figure}[h]
\label{fig2}
\begin{center}
\includegraphics[width=3in]{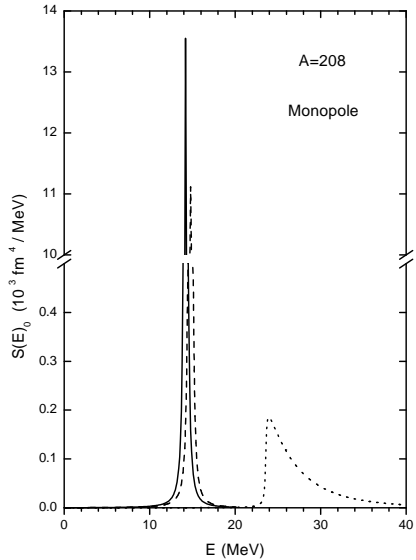}
\caption{Monopole strength function in three different approximations: zero-order (dotted curve), collective with finite value of $\kappa_{L}$ (solid) and collective with $\kappa_L=0$ (dashed). Note the change in the vertical scale.}
\end{center}
\end{figure}

For $L=0$ this expression gives the collective moving-surface strength function shown in Fig. 2 (full curve, $E=\hbar\omega$). 

The collective strength function of Fig. 2 has been calculated by assuming a residual interaction
\beq
\label{sepi0}
v_{L=0}(r,r')=\kappa_{L=0}r^2r'^2\,,
\eeq
with a value of $\kappa_{L=0} =-2\times10^{-4}\: {\rm MeV fm}^{-4}$. This parameter has been determined by fitting the experimental position of the giant monopole resonance in $^{208}$Pb. 

The dotted curve instead shows the zero-order strength function (proportional to the imaginary part of the response function ${\cal R}^0_L(\omega)$, which is similar to the quantum single-particle response function).

Finally, the dashed curve shows the collective moving-surface response given by Eq. (\ref{cmmsresp}) with $L=0$ and $\kappa_{L}=0$.  If $\kappa_{L}=0$, the frequency of the collective monopole vibration is determined by the solution of the equation
\beq
C_L-\chi_L(\omega)=0\,.
\eeq
It has already been pointed out in \cite{ads} that this approximation gives a very reasonable description
of the position (including the $A$-dependence) of the isoscalar giant monopole resonance in heavy nuclei.

Another interesting feature of the monopole response pointed out in \cite{ads} is that the zero-order  strength function vanishes for $\omega<\pi v_F/R$.  As a consequence of this fact, within this model, there is no Landau damping of the collective monopole mode. This absence of Landau damping is in qualitative agreement with the results of analogous quantum calculations \cite{dum, bertsch}.
The very small width appearing in Fig. 2 is due to our use of a fnite value of the infinitesimal parameter  $\varepsilon$ ( for numerical reasons, we have used $\varepsilon=0.1$ MeV).

We have checked numerically that the collective state shown in Fig. 2 exhausts about 99\% of the monopole energy-weighted sum rule, which is given by \cite{abr}
\beq
\label{ewsr0}
\int_{0}^{\infty} dE\,E\,S(E)=\frac{3}{10\pi}\frac{\hbar^{2}}{m}AR^{2}\,.
\eeq

\subsection{Dipole response (translation and compression modes)}

It is well known that the mean-field approximation violates the translation invariance of the nuclear Hamiltonian and that this results in the appearing of  spurious strength in the isoscalar dipole response. Hence the isoscalar dipole channel,  excited by the external field (\ref{mul}) with $L=1$, 
is usually not interesting because it should correspond to a simple translation of an unexcited nucleus, while in the mean-field approximation this pure translation is replaced by a spurious  excitation of the nuclear centre of mass bound by an unphysical force. However the corresponding compression mode, excited by the  field 
\beq
\label{mul2}
Q({\bf r})=r^{3} Y_{1M}({\bf \hat r})\,,
\eeq
has  received considerable attention because of the possibility of obtaining from it additional information about the compressibility of nuclear matter \cite{covg}. Since the external field (\ref{mul2}) can excite also the centre of mass, the problem of subtracting the unwanted spurious strength from the corresponding response function  has usually been dealt with by using an ingenious trick due to Van Giai and Sagawa \cite{vgs}. These authors suggested that, instead of studying the response to the external field (\ref{mul2}), one should look at the response to an effective external field of the kind
\beq
\label{effef}
Q_{eff}({\bf r })=(r^3-\eta r)Y_{1M}({\bf\hat r})\,,
\eeq
where $\eta$ is a parameter determined by the condition that, under the action of the external force, the  centre of mass should remain at rest. 

The moving-surface theory of \cite{ads} allows for a different approach to the problem of evaluating the intrinsic response associated with the field (\ref{mul2}). It is clear from the expression (\ref{rfp}) of the restoring-force parameter $C_L$ that this parameter vanishes for $L=1$. This means that in this case there is no restoring force, hence the moving-surface boundary condition seems to be able to readjust the translation symmetry that is broken by the mean-field approximation. This statement can be easily verified by looking at the form taken by the response function (\ref{cmmsresp}) when the residual interaction is neglected. If we put $\kappa_L=0$ in Eq. (\ref{cmmsresp}), the moving-surface response function becomes
\beq
\label{zomsresp}
{\tilde{\cal R}}_{L=1}^0(\omega)={\cal R}_1^0(\omega)
-\frac{[\chi^{0}_{1}(\omega)
]^{2}}{[-\chi_{1}(\omega)]}\,,
\eeq
or (cf. Eq. (3.1) of Ref. \cite{adm2})
\beq
{\tilde{\cal R}}_{L=1}^0(\omega)=\frac{3}{4\pi}\frac{A}{m\omega^2}\,,
\eeq
which has the behavior expected for a free particle. Since this response function has no poles for $\omega\neq 0$, it does not give spurious dissipation at positive $\omega$.

The translation invariance of the model when the residual interaction is taken into account is less obvious, however it has been shown in \cite{adm2},  by using sum rule arguments, that no  spurious strength is added to the zero-order intrinsic response  even when a residual dipole-dipole interaction is included. Thus we are confident that the present semiclassical theory with moving-surface boundary conditions correctly separates the intrinsic from the centre of mass excitations and do not need to use the effective operator (\ref{effef}).

For sake of simplicity, here we discuss in some detail only the zero-order response, in which the residual interaction is neglected. This approximation corresponds to treating the nucleus as a gas 
of non-interacting fermions confined to a spherical cavity with perfectly reflecting walls that are allowed to translate freely.
The residual interaction changes the compressibility of this nuclear fluid and its effects on the response have been evaluated in Ref. \cite{adm2}.

We first need to give a slight generalization of the zero-order response function (\ref{zor2}), by defining the functions
\beq
\label{zor3}
{\cal R}^0_{L,jk}(s)=\frac{9A}{16\pi}\frac{R^{j+k}}{\epsilon_F}\sum_{N=-L} ^LC_{LN}^2\sum_{n=-\infty}^{+\infty}
\int_0^1 dx x^2 s_{nN}(x) \frac{(Q_{nN}^{(j)}(x)/R^j)(Q_{nN}^{(k)}(x)/R^k)}{s-s_{nN}(x)+i\varepsilon}\,.
\eeq	
In terms of these new functions, 
the fixed-surface zero-order response to the external field (\ref{mul2}) is determined by the  function ${\cal R}^0_{1,33}(s)$, while  ${\cal R}^0_{1,13}(s)$ has a direct physical interpretation as the displacement of the nuclear centre of mass induced by the external field (\ref{mul2})  \cite{adm2}. Similarly  ${\cal R}^0_{1,11}(s)$ gives the response to the field $rY_{1M}({\bf \hat r})$.

\begin{figure}[h]
\label{fig3}
\begin{center}
\includegraphics[width=2in]{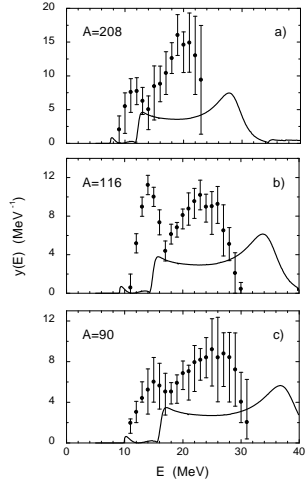}
\caption{ Comparison of our energy-weighted strength function with data from Ref. \cite{cla}. The curve shows the strength function in the case of vanishing residual interaction, i. e. for a confined Fermi gas with an incompressibility of $K=200$ MeV.} 
\end{center}
\end{figure}

It has been shown in \cite{adm2} that the (zero-order) moving-surface response function for the field (\ref{mul2}) can be written as
\beq
\label {a19}
\tilde {\cal R}_{1,33}^{0}(s)=\tilde {\cal R}_{c.m.}^{0}(s)+
\tilde {\cal R}_{intr}^{0}(s)\,,
\eeq
with
\beq
\label{a20}
\tilde {\cal
R}_{c.m.}^{0}(s)=\frac{3A}{4\pi}\frac{R^{6}}{2\epsilon_{F}}\frac{1}{s^{2}}
\eeq
and
\beq
\label{main}
\tilde {\cal R}_{intr}^{0}(s)={\cal R}_{1,33}^{0}(s)
-\frac{3A}{4\pi}\frac{R^{6}}{2\epsilon_{F}}\frac{1}{s^{2}}
{\Big \{}1-{{{\Big [}1-\frac{1}{2}s^2
\frac{{\cal R}_{1,13}^{0}(s)}{{\cal M}_{13}^{1}}
{\Big]}^2}\over{1-\frac{1}{2}s^2
\frac{{\cal R}_{1,11}^{0}(s)}{{\cal M}_{11}^{1}}}}
{\Big\}}\,.
\eeq

The moments
\beq
\label{mom}
{\cal M}_{jk}^{p}=\int_{0}^{\infty}dss^{p}{\big [}-\frac{1}{\pi}{\rm
Im}{\cal R}_{1,jk}^{0}(s){\big ]}
\eeq
are defined in terms of the generalized fixed-surface response functions  (\ref{zor3}) and they
can be easily evaluated, giving
${\cal M}_{11}^{1}=\frac{1}{3}\frac{9A}{16\pi}\frac{R^{2}}{\epsilon_{F}}$  and
${\cal M}_{13}^{1}=R^{2}{\cal M}_{11}^{1}$. 
An essential property of the intrinsic response function (\ref{main}) is
that its limit for $s\to 0$ is finite, so it has no pole in $\omega=0$.

A properly normalized and energy-weighted strength function associated with the intrinsic response function (\ref{main}) is shown in Fig. 3. It is interesting to note that the simple model used here qualitatively reproduces the  experimental data of \cite{cla}, in particular the double-peak structure of the dipole compression mode. Clearly at this level we can only hope in a qualitative agreement since the present model has several unrealistic aspects (sharp surface, no residual interaction, etc.).

\subsection{Quadrupole response}
\begin{figure}[h]
\label{fig4}
\begin{center}
\includegraphics[width=3in]{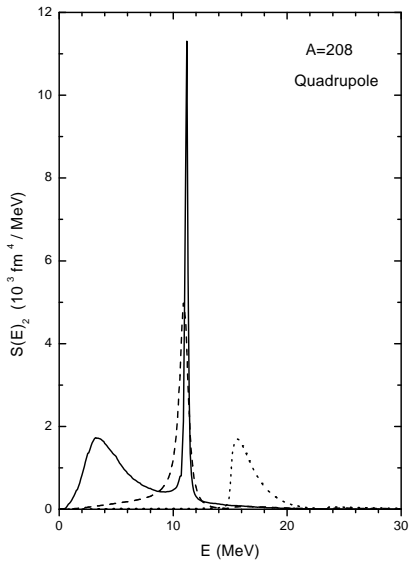}
\caption{Quadrupole strength function for a hypothetical nucleus of $A=208$ nucleons. The solid curve shows the moving-surface response, while the dashed curve gives the fixed-surface response. The dotted curve shows the response in the zero-order approximation. }
\end{center}
\end{figure}
Figure 4 shows the quadrupole response function given by Eq. (\ref{msresp}) (solid curve) with the strength of the residual quadrupole-quadrupole interaction determined by \cite{adm4}
\beq
\label{qqres}
\kappa_{L=2}=-1.0\times 10^{-3}\quad{\rm MeV\,fm}^{-4}\,.
\eeq

The value of this parameter has been fixed by the requirement that the  position of the giant quadrupole resonance (GQR) in our hypothetical nucleus of $A=208$ nucleons agrees with the experimental position of the GQR in $^{208}$Pb. The obtained value turns out to be about twice that given by the Bohr-Mottelson prescription (\cite{bm2}, p.509). Taking into account the fact that our equilibrium mean field has a different shape (square well rather than harmonic oscillator) and that we are assuming a semiclassical framework, this kind of agreement looks reasonable.  While the position of the GQR is well reproduced by the appropriate value of $\kappa_2$, its width is severely underestimated by our theory. This is a well known limit of all mean-field calculations in which the width is generated only by Landau damping, including a collision term in our kinetic equation would increase the width of this resonance. It is interesting to compare the moving surface response with the fixed-surface one (dashed curve in Fig. 4): contrary to the fixed-surface response function, the moving surface response displays a low-energy bump whose exact position is determined by the value of the surface tension parameter $\sigma$ of Eq. (\ref{rfp}). Thus the moving-surface theory does reproduce both systematic features of the quadrupole nuclear response that are the GQR and the low-energy surface excitations. Quadrupole response functions calculated for other values of $A$, corresponding to other medium-heavy spherical nuclei, are qualitatively similar to the case shown in Fig. 4.
\newpage
\subsection{Octupole response}

For $L=3$, Eq. (\ref{msresp}) gives the moving-surface collective octupole response function. A detailed study of this case has been made in Ref. \cite{adm5}. It is interesting to look first at the fixed-surface zero-order response. 
\begin{figure}[h]
\label{fig5}
\begin{center}
\includegraphics[width=2in]{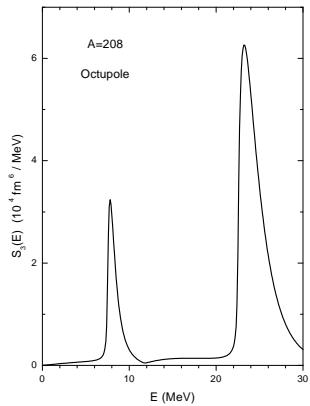}
\caption{Semiclassical octupole strength function analogous to quantum single-particle stregth function. Calculations are for $A=208$ nucleons in a square-well potentialof radius $R=1.2 A^{1/3}$ fm. }
\end{center}
\end{figure}
It can be seen from Fig. 5 that, for $A=208$, the single-particle octupole strength is concentrated
in two regions around 8 and 24 MeV. As pointed out already in \cite{bri}, in this 
respect our semiclassical response is strikingly similar to the quantum 
response, which is concentrated in the $1\hbar\omega$ and 
$3\hbar\omega$ regions. This concentration of strength is quite remarkable 
because our static distribution, which is given by Eq. (\ref{eqd}),
does not include any shell effect, however, because of the close 
connection between shell structure and classical trajectories 
expressed by Eq. (\ref{ediff}), we still obtain a 
strength distribution that is very similar to the quantum one. 

The collective moving-surface octupole strength function is shown instead in Fig. 6
(solid curve).
\begin{figure}[h]
\label{fig6}
\begin{center}
\includegraphics[width=3in]{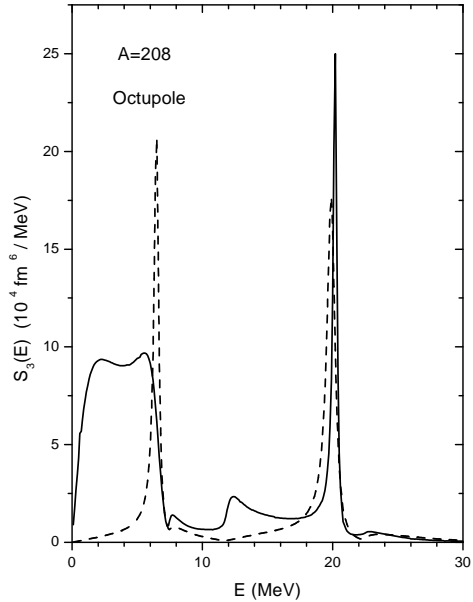}
\caption{The solid curve shows the octupole strength function given by the moving-surface solution (\ref{msresp}), while the dashed curve gives the corresponding fixed-surface response.}
\end{center}
\end{figure}

 Again we obtain a qualitative agreement with experiment and with the result of analogous quantum calculations. Like for the quadrupole case, agreement with experiment can be obtained with a residual interaction parameter about twice that given by the Bohr-Mottelson prescription. 
The rather broad double hump on the low-enegy side has been interpreted as a superposition of surface vibrations and of the low-enegy component of the giant octupole resonance. Within the present semiclassical theory, it can be shown that, for $L=3$, the parameters
$\delta R_{LM}(t)$, describing the octupole surface vibrations, approximately satisfy an equation of motion of the damped-oscillator kind \cite{adm5}:
\beq
\label{dosc}
D_3 \delta \ddot R_{3M}(t) +\gamma_3 \delta \dot R_{3M}(t)+C_3 \delta  R_{3M}(t) =0\,.
\eeq

The coefficients $D_3$, $\gamma_3$ and $C_3$ are easily evaluated, showing that the octupole surface oscillations described by Eq. (\ref{dosc}) are overdamped. Another interesting result of this semiclassical analysis concerns the shape stability of heavy spherical nuclei against octupole-type deformation: within the present model the spherical shape is stable.

 \section{Conclusions}

The linearized Vlasov equation, that can be seen as a particular  case of the Landau kinetic equation for the phase-space-density fluctuations,
gives a good qualitative description of the low-energy isoscalar nuclear response of different multipolarities.
This collisionless equation, which has been initially derived for other systems,  can be applied also to nuclei because in nuclear matter the mean-free-path of nucleons close to the Fermi surface is larger than the typical nuclear dimensions.  This fact has two consequences:
\begin{itemize}
\item{ the mean-field approximation is a reasonable one in the study of low-energy nuclear response,}
\item{ finite size effects are important and should be taken into account.}
\end{itemize}
Hence, because of the interplay between nucleon mean-free-path and nuclear dimensions, the Vlasov equation can be used to study the nuclear response to a weak driving field of long wavelength. Clearly in different physical situations, like those realized in the collisions of heavy ions of intermediate or high energy, collisions become more important and should be taken into account. Thus in nuclei finite-size effects are more important at low energy, while collisions between nucleons become more and more important with increasing energy.

As pointed out by Kirzhnitz and collaborators \cite{kir}, in finite systems, the boundary conditions satisfied by the fluctuations of the phase-space-density become essential. While in quantum mechanics these boundary conditions are authomatically enforced by the requirement that the wave function of bound states decreases exponentially outside the system, in the semiclassical kinetic-equation approach there is no similar requirement and the boundary conditions satisfied by the phase-space-density fluctuations must be imposed by using some reasonable criterion. Here we have studied the small fluctuations of the phase-space-density induced by applying a weak external driving force to spherical nuclei and have assumed a sharp-surface model for the spatial density and mean field in heavy nuclei.
Then we have compared the experimental isoscalar strength functions with those calculated by imposing two different kinds of boundary conditions on the phase-space-density fluctuations: fixed- and moving-surface boundary conditions. On the whole we find a better qualitative agreement with experiment for the moving-surface response. 

As a final comment, we would like to add that the present semiclassical theory can be applied also to other systems of many fermions in which it is important to take into account the finite size. Atomic clusters and magnetically trapped droplets of fermions are interesting examples. Clearly, quantum calculations are more appropriate, however it might be of some interest to see to what an extent the quantum features stand on a classical "skeleton", and this can be appreciated most clearly within the present approach.
\newpage

 \section*{Appendix A: Moving-surface response function}

In this Appendix we give a few more details on the derivation of the
analytical expression (\ref{msresp})
  for the multipole response function.

The  solution $\delta \tilde{f}^{L\pm}_{MN}$  of the linearized
Vlasov equation with the boundary condition (\ref{bc3}) can be
written as [cf. Eq. (\ref{sol1})]
\beq
\label{sol1ms}
\delta \tilde{f}^{L\pm}_{MN}(\epsilon,\lambda,r,\omega)=e^{\pm
i\phi_{N}(r,\omega)}[
\int_{r_{1}}^{r}d r'{ \tilde{B}}^{L\pm}_{MN}(r')
e^{\mp i\phi_{N}(r',\omega)]}\:+\tilde{C}_{\pm}(\epsilon,\lambda,\omega)]\,
\eeq
with the functions ${ \tilde{B}}^{L\pm}_{MN}$ given by an equation similar
to
(\ref{bfu})
\beq
\label{bfums}
\tilde{B}^{L\pm}_{MN}(\epsilon,\lambda,r,\omega)=
F'(\epsilon)[\frac{\partial}{\partial r}\pm \frac{iN}
{v_r(\epsilon,\lambda,r)}\frac{\lambda}{mr^2}][\beta Q_{LM}(r)+
\delta \tilde{V}^{int}_{LM}(r,\omega)]\,
\eeq
and the functions $\tilde{C}_{\pm}$ given by
  \beq
\label{msconst}
\tilde C_{\pm}(\epsilon,\lambda,\omega)=
\frac{e^{2i\phi_{N}(R,\omega)} {\tilde D}_{+}-{\tilde
D}_{-}}{1-e^{2i\phi_{N}(R,\omega)}}
+ F'(\epsilon)\frac{1}{\sin[\phi_{N}(R,\omega)]} mv_r(\epsilon,\lambda,R)
\omega\delta R_{LM}(\omega)\,.
\eeq

  The mean-field fluctuation $\delta \tilde{V}^{int}_{LM}(r,\omega)$
is a crucial quantity in our calculations.
  Usually  phenomenological models that describe the physical
properties of the medium differ mainly in the assumptions made about
this term. Our present approach is no exception to this general rule
and Eq. (\ref{msresp}) has been derived by assuming that
  \beq
\label{delvsepms}
\delta \tilde{V}^{int}_{LM}(r,\omega)=\beta\kappa_Lr^L
{{\cal R}}^V_L(\omega)\,,
\eeq
with
\beq
\label{mscollv}
{{\cal R}}^V_L(\omega)=\frac{1}{\beta}
\int dr r^2 r^L\delta \tilde{\varrho}_{LM} (r,\omega)\,,
\eeq
  and
  \bea
\label{denlm}
&&\delta \tilde{\varrho}_{LM} (r,\omega)=
\frac{8\pi^{2}}{2L+1}\frac{1}{r^{2}}\sum_{N=-L}^{L}
{\Big |}Y_{LN}(\frac{\pi}{2},\frac{\pi}{2}){\Big |}^{2}\nonumber\\
&&\int d\epsilon \int d\lambda \frac{\lambda}{v_{r}(\epsilon,\lambda,r)}
{\Big [}\delta {\tilde f}^{L+}_{MN}(\epsilon,\lambda,r,\omega)+\delta
{\tilde f}^{L-}_{MN}(\epsilon,\lambda,r,\omega){\Big ]}\,.
\eea

  The function (\ref{mscollv}) is not the response function
(\ref{msresp}) because, as discussed in \cite {adm2}, in the
moving-surface case the response function should include an
additional term in order to take into account the shape changes, thus
a more satisfactory definition of the multipole response function in
the moving-surface case is (see also \cite{jen})
  \beq
  \label{msrespf}
{\tilde{\cal R}}_L(\omega)=\frac{1}{\beta}
\int dr r^2 r^L\delta \bar{\varrho}_{LM} (r,\omega)\,,
\eeq
  with
\beq
\label{mden}
\delta \bar{\varrho}_{LM} (r,\omega)=
\delta {\tilde \varrho}_{LM}(r,\omega) +\varrho_{0}
\delta (r-R) \delta R_{LM}(\omega)\,,
\eeq
giving
\beq
  \label{msrespf2}
{\tilde{\cal R}}_L(\omega)={{\cal
R}}^V_L(\omega)+\frac{1}{\beta}R^{L+2}\varrho_0 \delta
R_{LM}(\omega)\,.
\eeq
The response function (\ref{msresp}) corresponds to (\ref{msrespf2}),
rather than to (\ref{mscollv}).
The equilibrium density $\varrho_{0}$ appearing in Eq. (\ref{mden})
is $\varrho_{0}=\frac{2}{3\pi^2}(p_F/\hbar)^3\,.$

In order to obtain the explicit expression (\ref{msresp}) of  the response 
function (\ref{msrespf2}), we need the explicit expressions
of the function ${{\cal R}}^V_L(\omega)$ and of the collective coordinates 
$\delta R_{LM}(\omega)$. For deriving these quantities, the moving-surface solution
(\ref{sol1ms}) should be expressed in terms of ${{\cal R}}^V_L(\omega)$ and  of
$\delta R_{LM}(\omega)$. By replacing the mean-field fluctuation 
(\ref{delvsepms}) into the quantities  ${\tilde{B}}^{L\pm}_{MN}$ and 
$\tilde{C}_{\pm}$, given by Eqs. (\ref{bfums}) and (\ref{msconst}), and
by inserting the resulting expressions into Eq. (\ref{sol1ms}), the fluctuation $\delta \tilde{f}^{L\pm}_{MN}$
can be written as
\bea
\label{sol2ms}
&&\delta \tilde{f}^{L\pm}_{MN}(\epsilon,\lambda,r,\omega)=
\delta f^{0L\pm}_{MN}(\epsilon,\lambda,r,\omega)
[1\,+\,\kappa_L{{\cal R}}^V_L(\omega)]\nonumber\\
 &&+\,F'(\epsilon)\frac{e^{\pm i\Phi_N(r,\omega)}}{\sin[\Phi_N(R,\omega)]} 
mv_r(\epsilon,\lambda,R)\omega\delta R_{LM}(\omega)\,,
\eea
with the zero-order solution $\delta f^{0L\pm}_{MN}$  given by
Eq. (\ref{zos}). 
Now, by inserting the solution (\ref{sol2ms}) into  Eqs. (\ref{mscollv}) and (\ref{delrlm}),
we obtain a system of  algebraic equations for the functions
${{\cal R}}^V_L(\omega)$ and $\delta R_{LM}(\omega)$, that can be written as
\beq
\label{mscolv2}
{{\cal R}}^V_L(\omega)={\cal R}^0_L(\omega)
[1\,+\,\kappa_L{{\cal R}}^V_L(\omega)] - \frac{1}{\beta}
{\Big [}\chi^{0}_{L}(\omega)+\varrho_{0}R^{L+3}{\Big ]}
\frac{\delta R_{LM}(\omega)}{R}
\eeq
and
\bea
\label{delrlm3}
\delta R_{LM}(\omega)=&&\frac{1}{C_L}{\Big \{}\beta R \chi^{0}_{L}(\omega)[1\,+\,\kappa_L
{{\cal R}}^V_L(\omega)] \\
 &&+ \chi_{L}(\omega)\delta R_{LM}(\omega)\,+\,
\beta \kappa_L \varrho_{0} R^{L+4} {{\cal R}}^V_L(\omega){\Big \}}\,,\nonumber
\eea
with the functions ${\cal R}^0_L(\omega)$, $\chi^{0}_{L}(\omega)$ and 
$\chi_{L}(\omega)$ given by Eqs. (\ref{zor}), (\ref{chi0l}) and 
(\ref{chil}), respectively.
Solving the system (\ref{mscolv2} -\ref{delrlm3}) gives the explicit 
expressions of  ${{\cal R}}^V_L(\omega)$ and  $\delta R_{LM}(\omega)$, which read
\beq
\label{mscolv3}
{{\cal R}}^V_L(\omega)={\cal R}_L(\omega) - \frac{1}{\beta}
\frac{\chi^{0}_{L}(\omega)+\varrho_{0}R^{L+3}}
{1 - \kappa_L{\cal R}^0_L(\omega)}
\frac{\delta R_{LM}(\omega)}{R}
\eeq
and
\beq
\label{delrlm4}
\frac{\delta R_{LM}(\omega)}{R}=\beta 
\frac{\chi^{0}_{L}(\omega)+\kappa_L\varrho_{0}R^{L+3}{\cal R}^0_L(\omega)}
{[C_L - \chi_{L}(\omega)][1-\kappa_L{\cal R}^0_L(\omega)]+
\kappa_L[\chi^{0}_{L}(\omega)+\varrho_{0}R^{L+3}]^2}\,.
\eeq
The fixed-surface collective response function ${\cal R}_L(\omega)$ appearing in Eq. (\ref{mscolv3}) is
given by Eq. (\ref{sepre}). 
Finally, by inserting Eqs. (\ref{mscolv3}) and (\ref{delrlm4}) into the 
response function  (\ref{msrespf2}) and taking into account that 
$\varrho_{0}R^3=\frac{3}{4\pi}A\,$, we find Eq. (\ref{msresp}).

\section*{Appendix B: Fourier coefficients}

In this Appendix we collect the expressions of the integrals (\ref{fc}) needed here. Since the spherical harmonics in (\ref{sh}) vanish unless $N$ has the same parity al $L$, we only need the corresponding integrals. Normally we need the coefficients with $k=L$, however for compression modes we also need $k=L+2$.
The coefficients involved in the monopole  response are
\bea
&&L=0,\quad N=0\nonumber\\
&&Q_{n0}^{(0+2)}(x)=\frac{2}{T}\int_{r_1}^{R} d r \frac{r^2}{v_r(\epsilon_F,\lambda,r)}\cos[\phi_{nN}(r)]\\
&&=(-)^nR^2\frac{2}{s_{nN}^2(x)}\quad{\rm for}\quad n\neq 0\nonumber\\
&&=R^2(1-\frac{2}{3}x^2)\quad{\rm for}\quad n= 0\,,\nonumber
\eea
while for the quadrupole response they are:
\bea
&&L=2,\quad N=0\nonumber\\
&&{\rm same\; as\; monopole}\nonumber\,,
\eea
and
\bea
&&L=2,\quad N=\pm 2\nonumber\\
&&Q_{nN}^{(2)}(x)=(-)^nR^2\frac{2}{s_{nN}^2(x)}{\Big (}1+N\frac{\sqrt{1-x^2}}{s_{nN}(x)}{\Big )}\,.
\eea

	In the dipole case we need the coefficients with
\bea
&&L=1,\quad N=\pm 1\nonumber\\
&&Q_{nN}^{(1)}(x)=(-)^nR\frac{1}{s_{nN}^2(x)}
\eea
for  translation modes and
\beq
Q_{nN}^{(3)}(x)=
(-)^nR^3 \frac{3}{s_{nN}^2(x)}{\Big (}1+\frac{4}{3}N\frac{\sqrt{1-x^2}}{s_{nN}(x)}-\frac{2}{s_{nN}^2(x)}{\Big )}
\eeq
for compression modes.

Finally for the octupole response we need:
\bea
\label{octfou}
&&L=3,\quad N=\pm 1,\pm 3\nonumber\\
&&Q_{nN}^{(3)}(x)=(-)^nR^3\frac{3}{s_{nN}^2(x)}{\Big (}1+\frac{4}{3}N\frac{\sqrt{1-x^2}}{s_{nN}(x)}\nonumber\\
&&-\frac{2}{s_{nN}^2(x)}+4(|N|-1)\frac{1-x^2}{s_{nN}^2(x)}{\Big )}\,.
\eea
For a given nucleus, the integrals $Q^{(k)}_{nN}$ could depend on two variables: the nucleon energy $\epsilon_F$ and its angular momentum $\lambda$. For the square-well potential however, they display a scaling property and depend only on the variable $x$. Moreover their $A$-dependence factorizes because $Q^{(k)}_{nN}\propto R^k$. As a consequence  the $A$-dependence factorizes also in the zero-order propagator (\ref{zor2}), that takes the form of an $A$-dependent factor times a universal propagator.

\end{document}